# Effects of nonmagnetic metal additives in metallic antiferromagnets on exchange bias

Marian Fecioru-Morariu[1], Syed Rizwan Ali[1], Cristian Papusoi[2], Martin Sperlich[1] and Gernot Güntherodt[1]

[1]Physikalisches Institut (IIA), RWTH Aachen University, 52056 Aachen, Germany

[2]SPINTEC, CEA/CNRS, 38054 Grenoble Cedex 9, France

The effect of nonmagnetic metal additives in metallic antiferromagnets (AFMs) on the exchange bias (EB) has been investigated from a structural, magnetic and Monte Carlo simulation point of view in bilayers of $Co_{70}Fe_{30}/(Ir_{22}Mn_{78})_{1-x}Cu_x$. Dilution by Cu atoms throughout the volume of the AFM $Ir_{22}Mn_{78}$ give rise to an enhanced EB field ($H_{EB}$) for 5 K ≤ T ≤ 300 K. The thermoremanent magnetization of the AFM-only is also enhanced with dilution, shows qualitatively the same temperature dependence as $H_{EB}$ and is at the origin of the EB mechanism. Monte Carlo simulations based on a Heisenberg model and temperature dependent uncompensated AFM moments confirm well the experimental results.

The phenomenon known as exchange bias (EB) refers to the shift of the hysteresis loop along the magnetic field axis originating from the interfacial exchange coupling between a ferromagnet (FM) and an antiferromagnet (AFM) [1]. Promising technological applications and challenging physics stimulated the intense research activities in this field [2,3]. Despite such focused attention, many aspects of exchange bias and related properties are not yet fully understood. This refers also to the role of nonmagnetic metal additives (dilution) in metallic AFMs [4-6] with respect to the strength of the shift field or EB field ($H_{EB}$).

In insulating AFMs it has been shown both experimentally and by Monte Carlo simulations that it is possible to enhance $H_{EB}$ by replacing magnetic atoms in the AFM by nonmagnetic ones (dilution by substitutional defects) [7-10] or by structural defects (twin boundaries) [11] not at the interface but rather throughout the volume part of the AFM. The enhancement of $H_{EB}$ in exchange-biased insulating CoO was attributed to uncompensated AFM magnetic moments [12] using a domain state (DS) model [7,8], which is based on the physics of diluted antiferromagnets in an external magnetic field [13,14] and on the Imry-Ma argument for domain formation [15].

In EB systems consisting of metallic AFMs and metallic FMs it has been noticed empirically that metal additives (M) have a distinct influence on $H_{EB}$. For example in the EB system $(CrMn)_{1-x}M_x$ an enhancement of $H_{EB}$ up to an $x$-dependent maximum was obtained at 300 K for $Ir_{0.03}$, $Pd_{0.05}$, $Pt_{0.08}$, $Cu_{0.08}$ and $Rh_{0.11}$ [4,5]. Also in the system NiFe/(PdMn)$_{1-x}$Pt$_x$ it was shown that $H_{EB}$ can be enhanced by increasing $x$ up to 0.3 [6]. No explanation of the $H_{EB}$ enhancement for these metallic AFMs was given so far.

In this Letter we analyze the effect of substitutional nonmagnetic metal defects in metallic exchange-biased AFMs on the enhancement of $H_{EB}$. As a model system we have chosen Cu as metallic additive in the metallic intermediate-anisotropy AFM IrMn exchange biased to CoFe. Lattice matched Cu dilutions give rise to a reduction of the AFM grain size and hence of the blocking temperature $T_B$ at which $H_{EB}$ vanishes. The enhancement and the maximum of $H_{EB}$ as a function of Cu dilution are found to result from the competition of the increased number of uncompensated AFM moments or DS magnetization ($M_{DS}$) and the reduction of $T_B$. The dilution and temperature dependence of $H_{EB}$ is found to have its origin in the thermoremanent magnetization ($M_{TRM}$) of the AFM itself. The temperature and dilution dependence of $H_{EB}$ is corroborated by Monte Carlo simulations considering a granular AFM structure and temperature dependent uncompensated AFM moments.

A series of samples with the layer sequence, Si/SiO$_2$/Cu (15 nm)/Co$_{70}$Fe$_{30}$ (6 nm)/ (Ir$_{22}$Mn$_{78}$)$_{1-x}$Cu$_x$(10 nm)/Au(2.5 nm), (denoted by series A in the following), was deposited by molecular beam epitaxy. The AFM of our investigation, Ir$_{22}$Mn$_{78}$, is a metallic AFM ($T_N$ = 550 °C) with an intermediate anisotropy constant of $K = 2 \times 10^6$ erg/cm$^3$. Within the DS model the dependence of $H_{EB}$ on anisotropy yielded a maximum at intermediate anisotropy values of the AFM [16]. As a suitable nonmagnetic impurity we chose Cu, since its lattice constant ($a_{Cu}$=3.615 Å) is close to that of the IrMn ($a_{IrMn}$= 3.71 Å), both lattices being fcc. Magnetic characterizations were performed by SQUID magnetometry after 1 T field cooling. The switching fields of the hysteresis cycles $H_1$, $H_2$ were used to determine $H_{EB}$ according to $H_{EB} = (H_1+H_2)/2$.

All the samples studied in the present work have a granular structure as for example shown in Fig. 1, where the atomic-force microscopy scans for samples with two different Cu dilutions are presented. We observed that with increasing Cu dilution in IrMn the average grain size decreases from about 65 nm for the undiluted sample to about 43 nm for the sample which contains 20 % Cu dilution in IrMn. This is in contrast to, e.g., metal additives in CrMn for which the grain size was assumed to be unchanged [4,5].

The temperature dependence of $H_{EB}$ for different Cu dilutions in IrMn is shown in Fig. 2(a). For all samples $H_{EB}$ was found to decrease with increasing temperature. This decrease becomes more abrupt at higher dilutions ($x > 10$ %), with $T_B$ shifting towards much lower values ($T_B \leq 100$ K). Furthermore, from Fig. 2(a), where for clarity only a few dependences are shown, it can be seen that $H_{EB}$ shows a maximum as a function of dilution. The effects of Cu dilution on $H_{EB}$ of all samples are summarized in Fig. 2(b). A distinct enhancement of $H_{EB}$ with dilution was observed for each temperature in the range from 5 K to 300 K. The maximum of $H_{EB}$ shifts with increasing temperature to smaller $x$ values. Beyond each maximum $H_{EB}$ drops for larger concentrations of Cu substitutional defects. At 300 K an enhancement of $H_{EB}$ of more than 60% is observed for a Cu dilution in IrMn of only 5% while $T_B$ has still a high value above 350 K.

In order to understand the origin of the above features of $H_{EB}$ and $T_B$ of our samples, sole AFM layers of the type Si/SiO$_2$/Pt(25nm)/Cu(10nm)/(Ir$_{22}$Mn$_{78}$)$_{1-x}$Cu$_x$(10nm)/Au(3nm), (denoted by series B in the following), were evaporated under the same deposition conditions as we used for series A. $M_{TRM}$ was measured by SQUID magnetometry, after the samples were field cooled in an external field of 7 T. A strikingly good qualitative agreement was observed by scaling the temperature dependence of $H_{EB}$ (of FM/AFM bilayers) with that of $M_{TRM}$ (of AFM only) as shown in Fig. 3 for the case of $x$ = 15 %. This observed similarity between $H_{EB}(T)$ and $M_{TRM}(T)$ clearly suggests that, for a given defect concentration in the AFM, a similar AFM domain and spin structure could be created either by the FM/AFM interaction or simply by a strong external field during cooling of the sole AFM. In the context of this discussion it is interesting to compare the dilution dependence of the $M_{TRM}$ of the sole AFM (inset of Fig. 3) with that of $H_{EB}$ (Fig. 2(b)). An evident resemblance in the behaviour of $M_{TRM}$ and $H_{EB}$ with increasing dilutions can be easily observed. However, at the same temperature the maximum of $M_{TRM}$ occurs at higher Cu dilution than the corresponding maximum of $H_{EB}$. This can be attributed to the contribution of both the surface and the volume AFM uncompensated moments to $M_{TRM}$, whereas $H_{EB}$ is determined by the surface part of the AFM uncompensated moments next to the adjacent FM layer. We have also compared the temperature at which $M_{TRM}$ of the AFM-only vanishes ($T_b$) with that at which $H_{EB}$ of the FM/AFM bilayer vanishes ($T_B$). We observed a similar behaviour with close values of $T_b$ and $T_B$ as a function of Cu dilution (not shown). This is in support of the fact that it is the $M_{TRM}$ of the AFM which governs the EB in the system under consideration.

In order to better understand the above observations, Monte Carlo (MC) simulations were performed for a FM/AFM bilayer using the DS model [7,8]. A DS throughout the volume of the AFM is stabilized when the diluted AFM is cooled below the Néel temperature in an external magnetic field. In this context, the $M_{DS}$ is related to the $M_{TRM}$ measured for series B. Nonmagnetic defects stabilize energetically the domain walls [8,14] and give rise to a surplus $M_{DS}$, the surface part of which interacts with the adjacent FM [8]. The irreversible part of this exchange interaction is responsible for the EB. For a realistic simulation of the CoFe/IrMn bilayers we consider their granular structure and assume temperature dependent uncompensated moments of the AFM grains through an Arrhenius-Neel law. In order to simulate an AFM with a concentration of substitutional defects, $x$, a fraction $x$ of randomly selected sites in the AFM was left unoccupied, while the residual sites, 1-$x$, carry a magnetic moment. In the simulations we used a time-quantified MC algorithm which is based on a Heisenberg model and thus capable of describing the three dimensional rotation of the spin moments. More details about our model and the parameters used are published elsewhere [17].

Figure 4 shows the simulated EB field ($H_{eb}$) of an FM/AFM bilayer for different dilutions in the AFM as a function of temperature normalized by the AFM exchange constant $|J_{AFM}|$. The decrease of $H_{eb}$ with temperature is found to be due to a decrease of the $M_{DS}$ via thermally activated spin relaxations in the AFM. At very low temperature some of the frustrated (uncompensated) spins, mainly those having the lowest coordination number due to surrounding defects, start to undergo thermally activated switchings. This results in an abrupt loss of $M_{DS}$ and causes the initial drop of $H_{eb}$. At higher temperatures the relaxation is relatively gradual, because it involves the uncompensated AFM moments which are strongly exchange coupled within AFM domains. Comparing the three curves in Fig. 4, an enhancement of $H_{eb}$ is obtained by increasing the dilution from $x = 30$ % to $x = 50$ % for $k_BT/|J_{AFM}| < 1.0$ (T < 100 K in the experiment; see inset).

The dilution dependence of $H_{eb}$ exhibiting a maximum is determined by the competition between the enhanced $M_{DS}$ due an increasing number of uncompensated AFM moments and the decrease of the blocking temperature. We have also observed that the decrease of blocking temperature with increasing dilution is due to the contribution of two factors: the decrease of the AFM grain size (observed experimentally (Fig. 1) as well as confirmed by Monte Carlo simulations [17]), and the decrease of thermal stability of the AFM due to the loss of connectivities in the AFM lattice upon dilution. At higher dilutions the increasing number of missing AFM bonds causes the observed decrease in the simulated blocking temperature and in $H_{eb}$ at, e.g., $x = 70$ %. The experimental temperature dependence of $H_{EB}$ of CoFe/(IrMn)$_{1-x}$Cu$_x$ bilayers for some selected dilutions is shown in the inset of Fig. 4. A very good qualitative agreement of the temperature and dilution dependence of the exchange bias field and the blocking temperature between the experimental and the simulated data is clearly visible. It is important to note that the defect concentrations for the experimental curves are systematically smaller than the simulated ones. Such differences are commonly found between simulated and experimental results [7,8,17,18] and are due to a residual or background defect concentration in the experimental samples for x = 0 [11].

In the system under consideration, we exclude the fact that the dilution dependence of $H_{EB}$ and $T_B$ could arise from an increase in the nearest neighbour Mn-Mn atomic distance. This was suggested by Soeya *et al.* [4] for explaining the enhancement of $T_B$ of CrMn with metal additives. In their case, a lattice mismatch of 35% appears by inserting the fcc Pt nonmagnetic defects into the bcc CrMn. Such a strain due to lattice mismatch could be responsible for the increase of the nearest neighbour Mn-Mn atomic distance and might give rise to the enhanced $H_{EB}$. However, in the EB system we have studied, the lattice mismatch between the fcc Cu nonmagnetic defects and the fcc AFM IrMn is only 2.5%. Therefore we can assume that the Cu atoms in IrMn give rise to a negligible strain in the host lattice [5,17].

In conclusion, the present investigation has shown that the EB of the bilayer CoFe/(IrMn)$_{1-x}$Cu$_x$ can be enhanced significantly due to Cu dilution in the AFM IrMn. The underlying origin is identified by the close qualitative connection between the temperature and dilution dependence of the $M_{TRM}$ of the AFM-only (series B) and that of the exchange bias field $H_{EB}$ of FM/AFM bilayers (series A). The temperature and dilution dependence of $H_{EB}$ is described by MC simulations based on the DS model, in which $M_{TRM}$ is related to the $M_{DS}$. Dilutions are found to increase $M_{DS}$ and to reduce the blocking temperature, explaining the enhancement and maximum of $H_{EB}$ for 5 K ≤ T ≤ 300 K. On the other hand, in the exchange-biased low-anisotropy AFM FeMn this is limited to 5 K only [17]. Our model-based explanation of the effect of metal additives in metallic AFMs on $H_{EB}$ also predicts a further increase of $H_{EB}$ if the grain size reduction upon dilution could be suppressed.

Two of the authors (M.F.M. and C.P.) would like to acknowledge the financial support received in the framework of the NEXBIAS Research Training Network (Contract No. HPRN-CT-2002-00296) financed by the EU. S.R.A. is grateful for funding by the Higher Education Commission, Government of Pakistan, under DAAD/HEC scholarship programme.

**FIG. 1.** Atomic-force microscopy surface scans of $(IrMn)_{1-x}Cu_x$ for two different Cu dilutions in IrMn (a) $x = 0$ and (b) $x = 0.2$. Scan size: x and y (in plane): 0.2 µm/div, z: (a) 25 nm/div (b) 10 nm/div

**FIG. 2.** Exchange bias field $H_{EB}$ of $CoFe/(IrMn)_{1-x}Cu_x$ (a) as a function of temperature for different Cu dilutions and (b) as a function of Cu dilution at different temperatures.

**FIG. 3.** Thermoremanent magnetization $M_{TRM}$ and exchange bias field $H_{EB}$ as a function of temperature for, respectively, an $(IrMn)_{1-x}Cu_x$ sample and a $CoFe/(IrMn)_{1-x}Cu_x$ bilayer sample with x=0.15 each. The inset shows the $M_{TRM}$ of $(IrMn)_{1-x}Cu_x$ as a function of dilution $x$ at different temperatures.

**FIG. 4.** Simulated temperature dependence of the exchange bias field $H_{eb}$ of a FM/AFM bilayer for different dilutions $x$ in the AFM. The inset shows the experimentally observed exchange bias field $H_{EB}(T)$ of $CoFe/(IrMn)_{1-x}Cu_x$ for different dilutions $x$.

Fig. 1

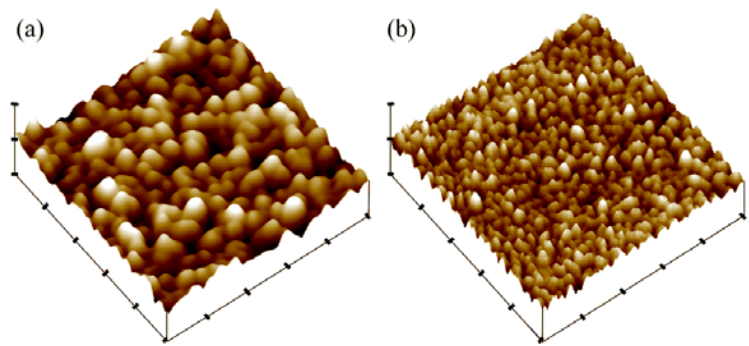

Fig. 2:

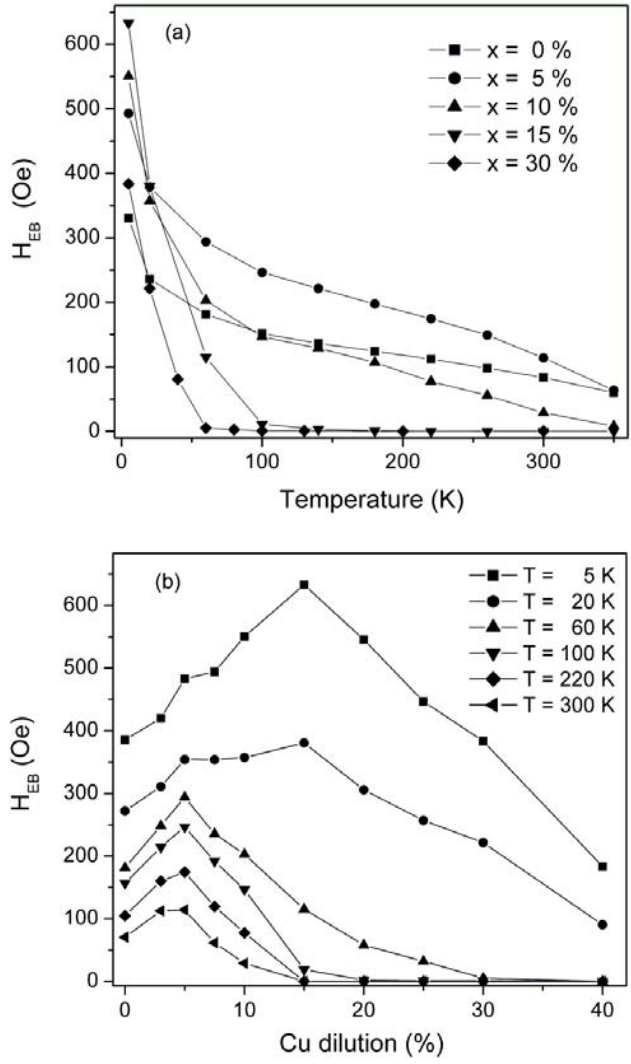

Fig. 3

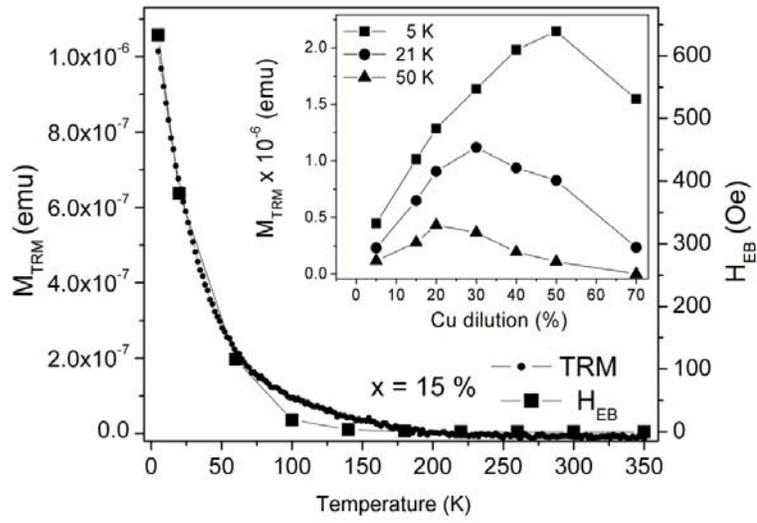

Fig. 4

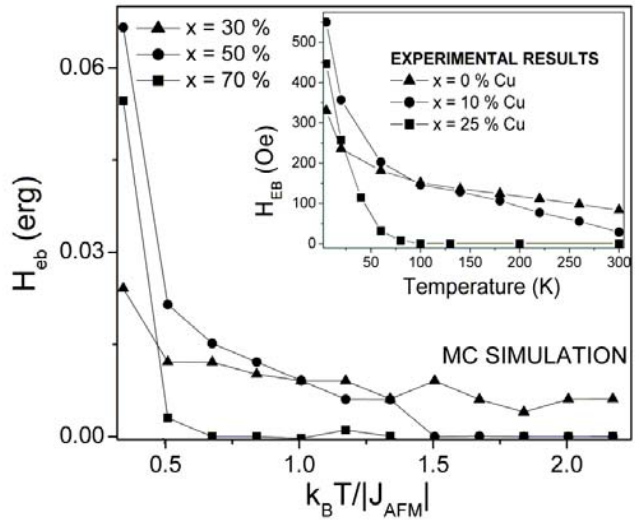